\documentclass[pra,superscriptaddress,showpacs,showkeys,twocolumn]{revtex4-1}

\usepackage{graphicx}
\usepackage[caption=false]{subfig}
\usepackage{placeins}
\usepackage{tikz}
\usepackage[normalem]{ulem}

\usepackage{amssymb}
\usepackage{amsmath}
\usepackage{soul}

\usepackage{physics}

\usepackage{ragged2e}

\newcommand{\om}{\omega}
\newcommand{\aop}[1]{\hat{a}_{#1}}
\newcommand{\tao}[1]{\hat{\widetilde{a}}_{#1}}
\newcommand{\co}[1]{\hat{a}_{#1}^{\dagger}}

\newcommand{\dt}[1]{\frac{\partial{#1}}{\partial t}}
\newcommand{\eps}[1]{\epsilon_{#1}}

\newcommand\wvsout{\bgroup\markoverwith{\textcolor{blue}{\rule[0.5ex]{2pt}{0.4pt}}}\ULon}

\DeclareMathOperator{\diag}{diag}

\usetikzlibrary{decorations.markings}
\tikzset{->-/.style={decoration={
  markings,
  mark=at position .5 with {\arrow{>}}},postaction={decorate}}}

\begin{document}

\title{Unitary work extraction from a Generalized Gibbs Ensemble using Bragg scattering}

\author{Wouter Verstraelen}
\address{TQC, Universiteit Antwerpen, B-2610 Antwerpen, Belgium}

\author{Dries Sels}
\address{TQC, Universiteit Antwerpen, B-2610 Antwerpen, Belgium}
\address{Department of Physics, Boston University, 590 Commonwealth Ave., Boston, MA 02215, USA}

\author{Michiel Wouters}
\address{TQC, Universiteit Antwerpen, B-2610 Antwerpen, Belgium}

\begin{abstract}

We investigate work extraction from integrable quantum systems under unitary operations. As a model system, we consider non-interacting fermions in one dimension. Thanks to its integrability, this system does not thermalize after a perturbation, even though it does reach a steady state which can be described by a Generalized Gibbs Ensemble (GGE). Such a GGE has an excess free energy compared to a thermal state and we propose to extract this energy by applying Bragg pulses.
We show how all the available work in the GGE can be extracted in the adiabatic limit while some excess energy is left at finite times. The unextracted work reaches the adiabatic limit as a power law with exponent $z=-2$ for small systems and with $z=-1$ in the thermodynamic limit. Two distinct protocols for combining the Bragg operations are compared, and in some systems an extensive difference in efficiency arises. From the unextracted work and the entropy production, a notion of temperature is defined and compared to the Boltzmann-Gibbs temperature of the system.
\end{abstract}

\maketitle

\section{Introduction} \label{sec:intro}

Over the last decade, an increasing amount of attention has been devoted to the interplay of thermodynamics, quantum mechanics and information theory \cite{Allahverdyan1,major,majorization,NatureHorodecki,naturework,atoommotor,therminfo,Allahverdyan2,Allahverdyan3,PhysRevLett.101.220402,Boksenbojm20104406,PhysRevE.94.010103,PhysRevE.90.042146,PhysRevX.6.041010,stochasticrealisation,PhysRevX.5.031044,PhysRevX.3.041003,beyondF,infotoEexp}. While most of the work is driven by fundamental questions regarding the validity of statistical mechanics in the quantum realm, recent technological advances have made it possible to actually fabricate and study thermal machines on the level of single-atoms. 
 In \cite{refereemachine1} a minimal universal heat machine is constructed with permanent coupling to heat baths; Ref. \cite{refereemachine2} adresses the distinction between work and heat in quantum machines and the quantum Otto cyclus is reviewed in \cite{kosottoreview}.

A prominent question is whether the ability to coherently control quantum systems allows one to construct engines that surpass the performance of their classical thermodynamic counterparts. 
There have been a number of approaches to this question, for example in \cite{coherence} it is shown that coherence itself can serve as a fuel such that work can be extracted from a single bath. In \cite{manycycles}, the authors account for temporal nonclassical correlations and how these are affected by measurements of total energy.
This type of questions is by no means new and was already at the hart of Maxwell's famous demon experiment. Since Landauer, it is well understood that any form of information should be considered as a thermodynamic resource. In that respect, it is the information about time \cite{clocks}, that allows one to use coherence or out-of-equibrium baths as resource. 

Here we consider a related but somewhat simpler problem, that does not involve time as the source of information. Imagine we have a system in equilibrium but in a non-thermal distribution. In particular, our system of interest will be an integrable quantum system, meaning it is non-ergodic due to an extensive number of conserved quantities. We will investigate how to use these conserved quantities as a resource to extract work from these systems. The crucial point is that due to the additional conservation laws, the steady state of the system has an entropy that is lower than the Gibbs entropy. Consequently, it should be possible to reduce the energy by an isentropic process, that brings the state to a lower energy Gibbs state.

In general, the linear structure of quantum mechanics makes all projections on eigenstates conserved, as opposed to classical ensembles where only the energy and particle number are conserved \cite{noneq.review}. It is thus far from trivial that in many cases, most of these conserved quantities are not relevant. A way to reconcile ergodic thermodynamics with this quantum picture is provided by the \emph{Eigenstate thermalization hypothesis(ETH)} \cite{ETHdeutsch,ETHsred,ETHnature,alessio}. In non-integrable systems the diagonal entropy usually becomes the Gibbs entropy. That means that without knowledge of the phase of the state, i.e. without coherence, one cannot extract work from these systems.

This does not imply, though, that conserved quantities are never important. When they are, the system is called \emph{integrable} and it does not thermalize.
As a more refined statistical description of such integrable quantum systems, one replaces the usual Boltzmann-Gibbs ensemble by a \emph{Generalized Gibbs ensemble (GGE)} of the form
\begin{equation}\label{eq:gge}
\hat{\rho}_{\text{GGE}}=Z^{-1} \exp\left(-\sum_i \eta_i \hat{I}_i \right),
\end{equation}
where each $\eta_j$ is a Lagrange multiplier that fixes the expectation value of $\hat{I}_j$. This GGE, as introduced by Jaynes \cite{jaynes1,jaynes2}, is readily obtained by constrained maximization of the von Neumann-entropy. In this context, it was first adopted by Rigol et al. \cite{rigol}. For completeness, we mention that in addition to integrability, thermalization can also be suppressed because of particular interference called many-body localization \cite{manybodyloc}, although the typical context in which the latter typically emerges differs from our set-up where we will work with momentum states. Finally, there also exists a transition regime between integrability and chaoticity in which the system relaxes to a non-thermal state, keeping memory of the initial state \cite{refereechaoticity1,refereechaoticity2}. 

An immediate consequence of integrability is that the entropy does not reach its maximal value at late times. The GGE has an entropy that is smaller than or equal to the Boltzmann-Gibbs entropy, where the equality is reached only when all Lagrange multipliers, apart from the inverse temperature $\beta$ are zero. Consequently, a GGE always has more free energy than a Gibbs state with the same entropy. This immediately raises the question whether this energy can be extracted to perform work. The answer is most certainly yes if one allows general isentropic manipulations. Coupling the system to an environment might however destroy the integrability of the system and we therefore restrict our analyses to unitary manipulations.

While all unitary processes are isentropic, not all isentropic operators are unitary. Apart from conserving the entropy, unitary operations conserve the full spectrum of the density matrix. As shown in \cite{Allahverdyan2}, it is always optimal in terms of work extraction to end in a state with no coherence. Consequently the work is maximized by organizing the occupations in descending order in terms of energy, i.e. the state becomes passive \cite{book:thirring}. Note that this can always be achieved by successive permutations of the occupation numbers. 

In the GGE description of an integrable quantum system, the Lagrange multipliers $\eta_j$ must be permuted to extract the maximal amount of work in such a way that they obtain the same ordering as the energy levels, corresponding to an inverse ordering of the occupation numbers. General considerations of work extraction from GGE states are also given in \cite{GGEwork,resource1,kossuggest2} ; work extraction from a resource-theoretical point of view has been investigated in Ref. \cite{resource2}.

As will be outlined in more detail below, our physical system of choice is a fermionic chain which is a commonly studied object in the context of integrability \cite{rigol,superHubbard,tqcGGE,chainquench}. It can be mapped onto a chain of hard-core bosons by a Jordan-Wigner transformation. A possible experimental realization are cold atoms trapped in an optical lattice. We will demonstrate that the proper unitary operations to perform a permutation of the Lagrange multipliers of this system can be attained by using a Bragg-hamiltonian. Bragg-spectroscopy \cite{BEC,opticsreview} is commonly used to probe the structure of cold atom systems, for example in the context of the Mott-insulator to superfluid phase transition \cite{braggham,BEC1,BEC2,BEC3,BEC4}. In essence, from interfering lasers an optical potential is created in the form of a travelling wave. This potential then exchanges an amount of momentum $q$ and energy $\omega$ with the system. 

The remainder of the paper is organized as follows. In Sec.~\ref{sec:setup} the setup of the system is given and we explain how we will describe work extraction from it. Sec.~\ref{sec:results} contains results regarding extracted work and entropy production as a function of the speed at which the operations take place. Additionally, some effective temperatures relevant to the extraction are introduced and studied. Finally, Sec.~\ref{sec:concl} concludes this work. We construct initial GGE states to extract work from using a quantum quench described in Appendix \ref{ap:quench}. While the main body focuses on a system obeying a quadratic dispersion relation, Appendix \ref{ap:tb} compares this with a tight-binding situation where, due to the particular shape of the dispersion relation, some additional peculiarities come into play.

\FloatBarrier
\section{Set-up}\label{sec:setup}

We consider a 1D lattice of spinless fermions (fermionic chain) with length $L$ and periodic boundary conditions. Two separate dispersion relations are studied: (i) a free fermion gas with $\hat{H}_{\text{gas}}=\sum_{k} \epsilon_k \co{k} \aop{k}$, where $\aop{k}(\co{k})$ are the annihilation (creation) operators in momentum space, $k=k_n=2n\pi/L$ and corresponding energy $\eps{k}= \frac{k^2}{2m}$ (we set $\hbar=1$ throughout the article);  (ii) a tight binding model is discussed in Appendix \ref{ap:tb}.

Treating the occupation of each $k$-mode as an independently conserved quantity, the GGE has the form
\begin{equation}\label{eq:ensemble}
\hat{\rho}_{GGE}=\bigotimes_{k=k_1}^{k_L}\hat{\rho}_k=\frac{1}{Z}\exp\left\{-\sum_{k=k_1}^{k_L}\eta_k\co{k}\aop{k}\right\}
\end{equation}
where each Lagrange multiplier $\eta_k$ fixes the expectation value $\langle\co{k}\aop{k}\rangle=n_k$. This GGE-ensemble, being the tensor product of the two-by-two density matrices of each mode, has an entropy which is the sum of the von Neumann-entropies of all states, namely
\begin{equation}\label{eq:entropy}S(\hat{\rho}_{GGE})=\sum_{k=k_1}^{k_L}S(\hat{\rho}_k)=\sum_k\left[\ln(1+e^{-\eta_k})+\frac{\eta_k}{e^{\eta_k}+1}\right].\end{equation}

Note that the GGE-ensemble description amounts to an effective Fermi-Dirac distribution for each $k$-mode, that is $n_k=[\exp{\eta_k}+1]^{-1}$ where $\eta_k$ can be thought of as having the form 
\begin{equation}\label{eq:etas}
\eta_k=(\eps{k}-\mu_k)/\mathcal{T}_k.
\end{equation} 
We have the freedom to set the chemical potential $\mu_k=\mu$, a constant independent of $k$ (see section \ref{subsec:temperatures}). $\mathcal{T}_k$ then corresponds to the effective temperature of mode $k$. 

\subsection{Bragg-operation \label{sec:BO}}
In order to perform the permutations of the single-particle occupations, we suggest to add successive Bragg-pulses~\cite{BEC,braggham,opticsreview} to the system, each pulse of the sequence provoking a transposition (swap) of the occupations of two levels. These Bragg Pulses are described by the additional time-dependent Hamiltonian
\begin{align}
\hat{H}_B=\frac{V_0}{2}\sum_k\left[e^{-i\omega t}\co{k+q}\aop{k}+e^{i\omega t}\co{k}\aop{k+q}\right].
\label{eq:HBragg}
\end{align}
Here, $q$ and $\omega$ are the wave vector and frequency transferred by the Bragg field to the atoms. $V_0$ is the strength of the Bragg potential, which is related to the intensity of the pulses (see e.g. \cite{BEC}). The appropriate values for these parameters, together with the pulse duration $T_{\text{swap}}$, will be discussed below.
The Heisenberg equations-of-motion for the annihilation operators with Hamiltonian $\hat{H}(t)=\hat{H}_0+\hat{H}_B(t)$ read
\begin{equation}
i\dt{\aop{k}(t)}=\epsilon_k\aop{k}+\frac{V_0}{2}e^{-i\omega t}\aop{k-q}+\frac{V_0}{2}e^{i\omega t}\aop{k+q}\label{eq:heisenberg}
\end{equation}
In order to eliminate the explicit time-dependence on the right-hand side of \eqref{eq:heisenberg}, we move to the rotating frame $\tao{k}=e^{i\om \frac{k}{q} t}\aop{k}$, leaving a linear equation for the vector $A$ of rotating annihilation operators
\begin{equation}\label{eq:matrixvgl}
i\dot{A}(t)=M A(t),
\end{equation} 
that describes simultateous Rabi-oscillations for all  pairs of levels whose wave vectors differ by $q$. It is evident that the oscillation between two levels $k_1$ and $k_2=k_1+q$ is resonant when $\omega=\epsilon_{k_2}-\epsilon_{k_1}$, allowing to select the swap (transposition in the terms of group theory which we will use \cite{marshallgroup}) between these two levels by properly fixing the Bragg parameters. For a two level problemn(a system with only two momentum modes), the calculation can be done explicitly \cite{sakurai}. It follows firstly that  given this resonance condition a full transposition of occupation is done in a time \begin{equation}\label{eq:swap}T_{\text{swap}}=\pi/V_0.\end{equation} Keeping this constraint in mind, only one free Bragg parameter remains. Secondly, the larger $T_{\text{swap}}$, the better the selectivity of a single transition, i.e. minimal unwanted side effects on other, non-resonant, energy levels. From this we deduce that most work can be extracted in the adiabatic limit.

With the help of Eq. \eqref{eq:matrixvgl}, the time-dependence of the GGE can be rewritten as
\begin{align}
\hat{\rho}_{\text{GGE}}(t)&=\frac{1}{Z}\exp\left\{-\sum_k \eta_k \co{k}(t)\aop{k}(t) \right\}\\
&=\frac{1}{Z}\exp\left\{-\sum_k \eta_k \tao{k}^{\dagger}(t)\tao{k}(t) \right\}\nonumber\\
&=Z^{-1}\exp\left\{-A^\dagger(t)\cdot B\cdot A(t)\right\}\nonumber\\
&=Z^{-1}\exp\left\{-A^\dagger(0)\cdot e^{iMt}\cdot B\cdot e^{-iMt}  \cdot A(0)\right\}\nonumber\\
&=Z^{-1}\exp\left\{-A^\dagger(0)\cdot B(t) \cdot A(0)\right\}\label{eq:evolution}
\end{align}
With $\diag[B(t)]=[\ldots \eta_{k-1}(t), \eta_{k}(t),\eta_{k+1}(t) \ldots]$.
By neglecting the off-diagonal elements of $B(t)$ we obtain
\begin{equation}
\hat{\rho}_{\text{GGE}}=\frac{1}{Z}\exp\left\{-\sum_k \eta_k(t) \co{k}\aop{k} \right\},
\label{eq:trunc}
\end{equation}
such that the time-dependence has been brought to the Lagrange multipliers. This approximation corresponds to letting the system equilibrate between successive Rabi pulses (i.e. off-diagonal elements will dephase).

\begin{figure}[h]
\centering
\subfloat[Protocol 1]{
\begin{tikzpicture}[scale=0.9]
\vspace{10cm}
\node at (0,0) (A) [label={A}] {};
\node at (1,0) (B) [label={B}] {};
\node at (2,0) (C) [label={C}] {};
\node at (3,0) (D) [label={D}] {};

\draw (A) circle (1pt) [fill]
      (B) circle (1pt) [fill]
      (C) circle (1pt) [fill]
      (D) circle (1pt) [fill];

\path [<->,bend right=20,very thick] (A) edge node [above] {1} (B);
\path [<->,bend right=55,very thick] (A) edge node [above] {2} (C);
\path [<->,bend right=90,very thick] (A) edge node [above] {3} (D);
\end{tikzpicture}}
\qquad
\subfloat[Protocol 2]{
\begin{tikzpicture}[scale=0.7]
\node at (0,0) (A) [label={A}] {};
\node at (1,0) (B) [label={B}] {};
\node at (2,0) (C) [label={C}] {};
\node at (3,0) (D) [label={D}] {};

\draw (A) circle (1pt) [fill]
      (B) circle (1pt) [fill]
      (C) circle (1pt) [fill]
      (D) circle (1pt) [fill];

\path [<->,bend right,very thick] (A) edge node [below] {3} (B);
\path [<->,bend right,very thick] (B) edge node [below] {2} (C);
\path [<->,bend right,very thick] (C) edge node [below] {1} (D);
\end{tikzpicture}}
\caption{ The two protocols for decomposing a cycle into transpositions that were considered, depicted for a four-state system where all states rotate one position to the right. Letters denote different states and the numbers denote the order in which the operations take place.}
\label{fig:protocollen}
\end{figure}
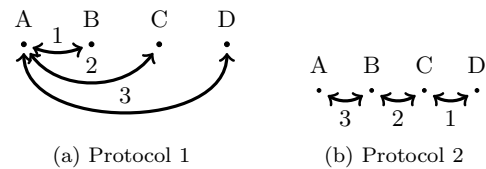

To extract all of the work, we will need to rearrange an extensive number of occupation numbers (or Lagrange multipliers). To execute the entire permutation with succesive Bragg pulses, we need to decompose the permuation in a series of transpositions. Recall that, any permutation can be decomposed in a cycle and each cycle of length $m$ can be written as $m-1$ successive transpositions \cite{marshallgroup}. The latter is non-unique, so we will compare two different protocols in this work, depicted in figure \ref{fig:protocollen}. Note that we can not perform the transpositions with unit fidelity because of the limited selectivity of the Bragg pulse.

In the case of free particles, a finite cutoff in k-space $k_{\text{cut}}$ has to be taken in order to keep the number of considered momentum modes finite. In the tight-binding case on the other hand, two new complications arise: periodicity in momentum space and coinciding resonances. The more complicate tight-binding case is discussed further in appendix \ref{ap:tb}. 

As an example, we apply our method to an initial non-thermal state created by a quench that constitutes instantly switching off an additional optical lattice potential. For details on this initial state, see appendix \ref{sec:quench}.

 \section{Results}\label{sec:results}

Starting from the initial state after suddenly turning off a periodic potential,
described in appendix \ref{sec:quench}, we examine the work that can be extracted by our Bragg procedure for both protocols. Work corresponds in this regard to the difference in the expectation value with respect to $\hat{\rho}_{\text{GGE}}$ of the energy between the state before ($\hat{\rho}_{\text{GGE},i}$) and after ($\hat{\rho}_{\text{GGE},f}$) execution of the whole protocol
\begin{equation}
W=\Tr[\hat{\rho}_{\text{GGE},f}\hat{H}_{\text{gas}}]-\Tr[\hat{\rho}_{\text{GGE},i}\hat{H}_{\text{gas}}].
\end{equation}
We define the \emph{maximal extractable work} $W_{\text{max}}$ as the work that a perfect permutation of the Lagrange multipliers would allow to extract, as described in the introduction. This quantity $W_{\text{max}}$ also goes under the name \emph{ergotropy} \cite{Allahverdyan2,refereemachine2}. The \emph{extracted work} $W$ on the other hand, corresponds to the work extracted with finite $T_{\text{swap}}$ using one of our Bragg protocols. The \emph{unextracted work} then is the difference between the \emph{maximal extractable work} and the \emph{extracted work} $W_{\text{unext}}=W_{\text{max}}-W$.
The \emph{entropy production} $\Delta S$ is the difference between the final and initial entropy of the state, where the entropy is given by \eqref{eq:entropy}.   We numerically implemented Eqs. \eqref{eq:evolution} and the dephasing \eqref{eq:trunc} for successive transpositions, obtained by an appropriate choice of the Bragg Hamiltonian \eqref{eq:HBragg}.
We have verified, by comparison, that ambiguity in the ordering when some modes are degenerate does not affect the results.

\subsection{Work extraction}

\begin{figure}[h!]
\includegraphics[width=\linewidth]{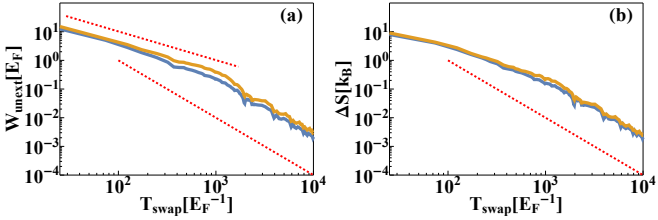}
\caption{Unextracted work and entropy production for protocol 1 (blue) and protocol 2 (yellow) as function of $T_{\text{swap}}$. $T_{\text{swap}}^{-2}$ and  $T_{\text{swap}}^{-1}$ lines (red, dashed) are shown for comparison.
Parameters $k_F L=50\pi$ and $k_{\text{cut}}/k_F$=2 were used.
}
\label{fig:kwadwork}
\end{figure}

Work extraction for both protocols is studied as a function of $T_{\text{swap}}$. The results are shown on fig. \ref{fig:kwadwork}. It is worth stressing that the work shown is the work for an entire process (as opposed to a single transposition) as function of the parameter $T_{\text{swap}}$ that characterizes the adiabaticity. First, protocol 1 turns out to be more efficient than protocol 2 as it needs a smaller $T_{\text{swap}}$ to extract the same amount of work. As expected, in the adiabatic limit all work can be extracted. Furthermore, both the unextracted work and the entropy production decay algebraically as $T_{\text{swap}}^{-2}$ in the long transposition time limit whereas the decay is slower for smaller $T_{\text{swap}}$, linear for $W_{\text{unext}}$.

Note that for short $T_{\text{swap}}$ a negative amount of total work $W=W_\text{max}-W_\text{unext}$ is extracted. For short $T_{\text{swap}}$ the Bragg operation is rather unselective; by coupling to too many unwanted states considerable entropy is generated. Since most of those states are at higher energy, the system is effectively being excited rather than cooled down.

The behavior of the unextracted work for long transposition times can be understood analytically. Since the values of both unextracted work and entropy production are small, we assume them to be linear in the number of unwanted (off-resonant) transpositions. Generalizing the expression for a single Rabi-oscillation (see for example \cite{sakurai}) to a superposition of all simultaneous oscillations, the total number of particles transferred to another mode after time $T_{\text{swap}}$ can be written as

\begin{equation} \label{eq:kwaddecay}
\sum_{k}\frac{\sin^2\left(\sqrt{1+\left(\frac{\eps{k+q}-\eps{k}-\om}{V_0}\right)^2}V_0 T_{\text{swap}}\right)}{1+\left(\frac{\eps{k+q}-\eps{k}-\om}{V_0}\right)^2}N_{k}
\end{equation}
where $N_{k}$ denotes the occupation of mode $k$ before the operation. We can easily extract an upper bound for this quantity by nothing that all $N_{k}<\max N_{k} \leq 1$. Furthermore using $\eps{k}=\frac{k^2}{2m}$, expression \eqref{eq:kwaddecay} can be bounded by

\begin{equation} \label{eq:som}
\sum_{k} \frac{1}{1+ \left(\frac{2kq+q^2-\om}{2 m V_0}\right)^2}.
\end{equation}

Now suppose the purpose of the operation is transposing the states $k=0$ en $k=\frac{2\pi}{L}$, such that parameters $q=\frac{2\pi}{L},\,\omega=\frac{4\pi^2}{2 m L}$ must be chosen. We extend the sum over k-modes up to infinity.
The total of all off-resonant (and hence unwanted) contributions is then proportional to

\begin{align}
&2\sum_{n=1}^{\infty} \frac{1}{1+ \left(\frac{4\pi^2 n}{m L^2 V_0}\right)^2}=-1+\frac{m L^2 V_0}{4\pi}\coth{\frac{m L^2 V_0}{4\pi}}\label{eq:ksom}.
\end{align}

In the adiabatic limit $\frac{m L^2 V_0}{4\pi} \ll 1$, this expression can be expanded as
\begin{equation}
\frac{1}{3}\left(\frac{mL^2V_0}{4\pi}\right)^2+\mathcal{O}\left[\left(\frac{m L^2 V_0}{4\pi}\right)^4\right]\label{eq:ontwikkeling}, 
\end{equation}
which is quadratically increasing in $V_0$ in leading order, or equivalently quadratically decaying in $T_{\text{swap}}=\pi/V_0$.

In the thermodynamic limit $\frac{m L^2 V_0}{4\pi} \gg 1$ on the other hand, expression \eqref{eq:ksom} reduces to
\begin{equation}
-1+\frac{m L^2 V_0}{4\pi},
\end{equation}
Which is only linearly decreasing with $T_{\text{swap}}.$

If, instead of $k=0$ and $k=\frac{2\pi}{L}$, one wishes to transpose another pair of modes, the results remain valid as we show in the following. The difference between the energy levels $\eps{k+q}$ and $\eps{k}$ is $(2q+q^2)/(2m)$ and for the rotated states 
\begin{equation}\label{eq:rotediff}
\widetilde{\epsilon}_{k+q}-\widetilde{\eps{k}}=(2kq+q^2)/(2m)-\om.
\end{equation}
$\om$ is chosen such that the pair of levels $k'$ en $k'+q$ becomes resonant, from which
\begin{equation}
\widetilde{\epsilon}_{k'+q}-\widetilde{\epsilon}_{k'}=0=2k'q+q^2-2m\om\Rightarrow 2m\om=2k'q+q^2.
\end{equation}
Substituting in \eqref{eq:rotediff} then yields
\begin{equation}
\widetilde{\epsilon}_{k+q}-\widetilde{\epsilon}_{k}=\frac{q}{m}(k-k'),
\end{equation}
which depends only linearly on $(k-k')$.

As a consequence, choosing $k'\neq 0$ merely shifts the terms in the summation \eqref{eq:ksom}. From \eqref{eq:som}, it is also readily seen that operations where $q\neq \frac{2\pi}{L}$ corresponds to a rescaled version of \eqref{eq:ontwikkeling}. From this, it is apparent that the imperfections of the whole procedure, consisting of a number of successive operations where each instance the imperfections decay quadratically (linearly) with  $T_{\text{swap}}$, will also be decaying quadratically (linearly) in $T_{\text{swap}}$. The linearisation of the work functional in the number of transferred particles then provides the behaviour in both the linearly and quadratically decaying regime. For the entropy functional (which is less linear by itself), only the quadratically decaying regime is recovered [see Fig. \ref{fig:kwadwork} (b)].

\subsection{Effective temperatures} \label{subsec:temperatures}

Usually, the notion of temperature refers to states that thermalize, but we wish to investigate here whether it can be meaningful to define a temperature for the integrable system under consideration. In particular, we define three different temperatures for our GGE-state: the Boltzmann-Gibbs temperature $\mathcal{T}_{BG}$, the average mode-temperature $\mathcal{T}_{\text{av}}$ and the extraction temperature $\mathcal{T}_{\text{ext}}$.

The Boltzmann-Gibbs temperature $\mathcal{T}_{BG}$ is simply the equilibrium temperature that a Boltzmann-Gibbs state (Fermi-Dirac distribution of k-modes) would have if this equilibrium state has the same total energy $E$ and particle number $N$ as the GGE-state. It hence satisfies
\begin{align}
\sum_k \frac{\eps{k}}{e^{(\eps{k}-\mu)/\mathcal{T}_{\text{BG}}}+1}&=E\\
\sum_k \frac{1}{e^{(\eps{k}-\mu)/\mathcal{T}_{\text{BG}}}+1}&=N,
\end{align}
where $\mu$ is the chemical potential. 

Secondly, we define $\mathcal{T}_\text{av}$ to be the average of the temperatures of the individual $k$-modes
\begin{equation}
\mathcal{T}_\text{av}:=\langle \mathcal{T}_k \rangle_k,
\end{equation}
where each $\mathcal{T}_k$ is calculated from \eqref{eq:etas}. For $\mu$, the Boltzmann-Gibbs value mentioned above is taken here. 

For our final notion of temperature, we start from the thermodynamic relationship that the temperature is the derivative of energy with respect to entropy. In our case however there is no unique correspondence between these two quantities as the configuration space is L-dimensional (or, keeping the particle number constant, L-1 dimensional), so that the temperature is dependent on direction along which a variation occurs $\mathcal{T}_{\text{dir}}=\frac{\delta E}{\delta S}$. To be specific, we consider this temperature for the final state with parameters $(E_f,S_f)$, where the variations correspond to a change in $T_{\text{swap}}$. This corresponds to 
\begin{equation}
\mathcal{T}_{\text{ext}}=\frac{\delta E_f}{\delta S_f}=\frac{W_{\text{unext}}(T_{\text{swap}})}{\Delta S (T_{\text{swap}})}.
\end{equation}
We are particularly interested in the behaviour for $T_{\text{swap}}$ sufficiently long so that the finite time-protocol can be regarded as resulting in a small perturbation $(\delta E_f,\delta S_f)$ from the adiabatic result $(E_f^{(0)},S_f^{0}=S_i)$

These three different notions of temperature are shown on figure \ref{fig:kwadtemperatures} (see also Fig. \ref{fig:tbtemperatures} in Appendix \ref{ap:tb} for the tight binding case).
We see that $\mathcal{T}_{\text{ext}}$ is the most protocol-dependent as it depends on the small difference between the attained state and the perfect state rather than the attained state as a whole. Also some oscillatory behaviour is present, which might diminish for larger system size. For $\mathcal{T}_{\text{av}}$ and $\mathcal{T}_{\text{BG}}$ on the other hand, it becomes increasingly difficult to distinguish both protocols. Due to the fundamentally different nature of the three temperatures, it is still remarkable that they differ less than a factor two.

\section{Conclusions and outlook} \label{sec:concl}

In this work, we demonstrated how work can be extracted from an integrable quantum system described by a GGE, using fermionic chains as an example. The work extraction is done by permuting the Lagrange multipliers and we provide a method to construct the necessary unitary operations from Bragg spectroscopy, that can be implemented with ultracold atomic gases.
Our method was explicitly demonstrated for two separate dispersion relations: a free gas and a tight-binding lattice. In both cases, we started from an initial stationary state constructed by a quenching procedure. 

In the case of a free gas, all possible work becomes extracted in the adiabatic limit. When performing the operations with finite speed, the decay of the unextracted work and of the entropy production goes

 deep in the adiabatic limit as $T_{\text{swap}}^{-2}$ in the slowness of the operations and scales quartically with system size as $L^4$. For faster operations (shorter $T_\text{swap}$) or equivalently, larger systems, the unextracted work goes only linear as $T_\text{swap}^{-1}$ and increases quadratically with the system size as $L^2$. Because in all cases the dependence on $L$ and $T_{\text{swap}}$ is trough $L^2 T_{\text{swap}}^{-1}$, one needs to scale $T_{\text{swap}}\propto L^2$ in order to keep the excess density (average imperfection per mode) constant. As also the total number of transpositions to perform is extensive in the system size, the total time of executing the protocol scales at $L^3$ under constant excess density.

The two protocols we compared produce similar results in most cases. There are cases however, where an extensive difference between the two protocols is evident, for example when significant next-nearest-neighbour hopping is present in a tight binding model (see  Appendix \ref{ap:tb}). Other protocols may be constructed as well (or using a different starting point for the transpositions within the same protocol ) and it will be worthwhile to verify the efficiency of a protocol before implementing it in practice.

From the unextracted work and the entropy production, we defined an effective temperature, which was of the order of the Boltzmann-Gibbs temperature of the system.

Note that through the Bragg hamiltonian the optical potential is treated as a classical field. This means that photonic uncertainty and entanglement between the system and the environment (as entanglement within the system) are not incorporated in the picture used. We expect the influence of these principles to our results to be rather limited \cite{mekhov}, but nonetheless it may be interesting to look how they change the picture precisely.

The methods are also applicable to bosonic systems that can be mapped to a fermionic one by a Jordan-Wigner transform. On the other hand, the fermionic particles could also have a spin, in which case the polarization of the Bragg laser would distinguish the spin components.
Our method can be extended to higher-dimensional systems as well and can be applied to generic band structures. 

\begin{figure}
\centering
\includegraphics[width=0.7\linewidth]{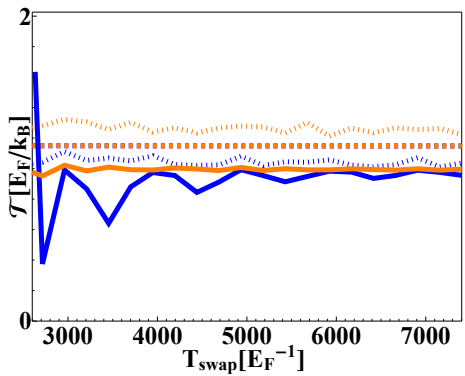}
\caption{ Comparison of the different measures of temperature as a function of $T_{\text{swap}}$ for the quadratic dispersion in  our $k_F L=50\pi,\, k_{\text{cut}}/k_F=2$ system. Dotted lines: $\mathcal{T}_{\text{ext}}$, dashed lines: $\mathcal{T}_{BG}$ and full lines: $\mathcal{T}_{\text{av}}$. The blue lines correspond to protocol 1 and the orange ones to protocol 2.}
\label{fig:kwadtemperatures}
\end{figure}

\section*{Acknowledgements}
We acknowledge discussion with J. Naudts. This work was financially supported by the FWO Odysseus program. D.S. acknowledges support of the FWO as post-doctoral fellow of the Research Foundation - Flanders.

\FloatBarrier
\appendix

\section{Constructing an initial GGE state by quenching}\label{ap:quench}\label{sec:quench}

In quantum mechanics, time-evolution is described by a unitary operation. On the other hand, changes of basis are also unitary transformations. The interplay of these two unitary operations can give rise to a number of different manifestations when both are present. In the case of adiabatic time-evolution, both operations coincide such that the density matrix after evolution expressed in the new basis has the same elements as the density matrix before the transformation, expressed in the old basis. The opposite limit, quenching, takes place when de hamiltonian is changed instantly~\cite{quenching}. Generally, the system is no longer in an eigenstate after this quench. However, as we only consider the occupations as conserved quantities and neglect the oscillating off-diagonal elements (decoherence which corresponds to time-averaging), the resulting density matrix becomes diagonal again, be it with different elements (eigenvalues). If one starts from a thermal state, the ordering of occupations is opposite to the ordering of the energy levels, as the Boltzmann-distribution is a decreasing function. After quenching, this order is generally no longer maintained. Because of this, quenching can be useful to construct a non-thermal stationary state from which work can be extracted.

\begin{figure}[h!]

\includegraphics[width=\linewidth]{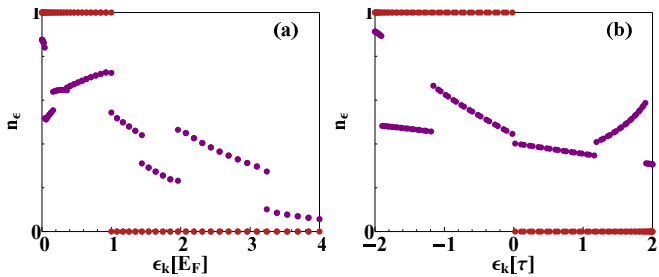}
\caption{Distribution of particles as function of energy before (red) and after (purple) a $J=8,\lambda=5,L=100$ quench. Left: quadratic dispersion and Right: ($\Delta k=0.01$-shifted) tight-binding dispersion. As the quadratic dispersion relation is symmetric, each dot corresponds to two opposite k-modes.}
\label{fig:distributions}
\end{figure}

In particular, we perform a quench consisting of switching off an additional term in the Hamiltonian of the form
\begin{equation}
\frac{J}{2}\sum_k (\co{k}\aop{k+\frac{2\pi}{\lambda}}+\co{k}\aop{k-\frac{2\pi}{\lambda}}),
\end{equation}
corresponding to an additional cosine-potential with period $\lambda$ (which is commensurable with $L$) and strength $J$. For the tight-binding case where periodicity has to be taken into account, the sums in $k$-labels are considered modulo $2\pi$. We rather consider the switching off than the switching on of the periodic potential in order to have the simpler final Hamiltonian. The state after the quench is then used as an initial state to extract work from. As particle distribution before the quench a zero-temperature Fermi-Dirac distribution at half filling is taken.
In the main body of the text, a quench with $\lambda=5, J=8$ and is used onto a system of 100 modes $(L=100$, largest considered wavenumber $k_{\text{cut}}=\pi)$ containing 50 fermions $(N=50,$ fermi-momentum $k_F=\frac{\pi}{2})$ The momentum-cutoff corresponds to a position resolution of $\frac{\pi}{k_{\text{cut}}}$ which, especially in the tight-binding case, can be thought of as the lattice distance of the system. The momentum $k_{\text{cut}}=\pi$ thus rather corresponds to the edge of the first Brillouin zone than a cutoff above which modes are not considered. This periodicity in $k$-space is discussed further in Appendix \ref{ap:tb}. The distributions of particles as function of the energy are shown on figure \ref{fig:distributions}.

\FloatBarrier

\section{tight-binding dispersion} \label{ap:tb}

In the tight-binding case, with Hamiltonian
\begin{equation}\hat{H}_{\text{tb}}
=-2\tau\sum_k \cos(k)\co{k}\aop{k}\end{equation}
some care must be taken. Because of the periodicity in momentum space, in the matrix $M$ from equation \eqref{eq:matrixvgl} an additional pair of nonvanishing (non-main) diagonals appear that correspond to the Umklapp processes that cross the boundary of the Brillouin zone. Contrary to the direct processes, the Umklapp matrix elements do not become time-independent under the transformation to rotating basis states. In order to still be able to perform an efficient computation where only diagonalization must be done numerically, we will therefore neglect these Umklapp-processes. However, for each Bragg operation we choose a new Brillouin zone, with a cut in k-space chosen such that the processes neglected are the transitions that are the least coupled with the resonant transition.

A second issue with the tight-binding dispersion 
is that the naive choice of basis states for $k$ as done above leads to multiple resonances that coincide. Namely, for certain combinations of $q$ and $\omega$ more than one pair of levels is at resonance, making it impossible to affect them independently. We propose two distinct solutions for this latter issue. The first possible solution is to use basis states that are slightly shifted in k-space, which corresponds to the presence of a gauge field~\cite{gaugefield}. An alternative solution is to introduce an additional term in the fermionic Hamiltonian, such as a next-nearest-neighbour hopping. Both solutions break the degeneracy of the Bragg transitions.

\begin{figure}[h!]
\centering

\includegraphics[width=\linewidth]{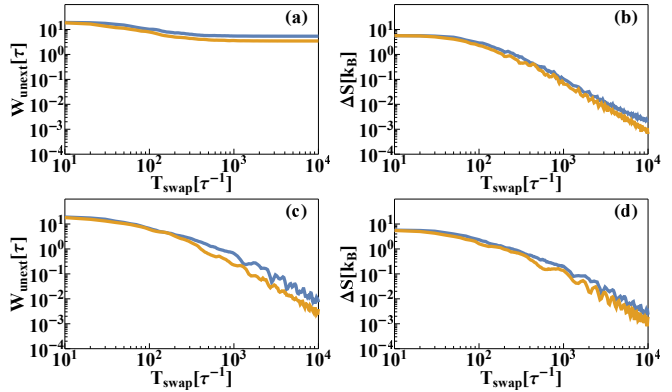}
\caption{Unextracted work (left) and entropy production (right) as function of adiabaticity for the unperturbed case (top) and assuming $\Delta k=0.01$ Galilei-shifted basis states (bottom). Results where the dispersion includes a $\sigma=0.01$ next-nearest-neighbour hopping term are numerically similar to the bottom case. Perturbing parameters for the latter two cases are chosen small to make comparison with the unperturbed case meaningful. Clearly, the maximal available amount of work $W_{max}\approx 5.19$ is reached in the adiabatic limit in the bottom case but not in the unperturbed one. Again protocol 1 is depicted in blue and protocol 2 in orange.}
\label{fig:tbwork}
\end{figure}

Results for $W_{\text{unext}}$ and $\Delta S$ are shown as function of $T_{\text{swap}}$ on figure \ref{fig:tbwork}. From (a),(b) it is seen that for the naive choice of basis states $k=2n\pi/L$ the extracted work does not approach the maximal extractable work in the adiabatic limit even though the entropy production tends to zero. This suggest a permutation of the Lagrange multipliers is performed which is not the optimal one, confirming the issue of coinciding resonances. Indeed, on (c), (d) it is shown that the addition of a small gauge field that slightly shifts the dispersion allows to extract all of the maximal extractable work. Similarly, also adding a small next-nearest-neighbour (nnn)-hopping perturbation to the Hamiltonian results in a full extraction of the available work. The behaviour in these two cases can be attributed to the possibility of distinguishing the transition degeneracies as discussed above.

In contrast to the case of the quadratic dispersion, there is no simple analytical expression that describes the decay as function of $T_{\text{swap}}$. One of the reasons for this is that the energy states are on average closer to each other here, making the number of significant off-diagonal processes larger and resulting in a decay that is slower than in the case of the quadratic dispersion. For both protocols nevertheless the unextracted work and the entropy production decay to zero in a similar way. As opposed to the case of the quadratic dispersion, protocol 2 seems more efficient on average although the difference between the two protocols remains modest.

\begin{figure}[h!]
\centering

\includegraphics[width=\linewidth]{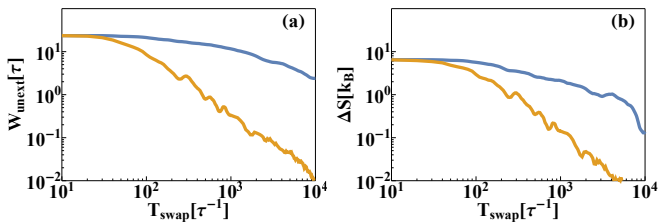}

\caption{Extracted work (a) and entropy production (b) for a tight-binding lattice including larger next-nearest neighbour $\sigma=0.1 \tau$ hopping for protocol 1 (blue) and protocol 2 (orange). This situation is an example where the choice of protocol will make a significant difference.}
\label{fig:largesigma}
\end{figure}

This is not always the case though, as is seen on figure \ref{fig:largesigma}. When there is a next-nearest-neighbour hopping term $\sigma=0.1\tau$ present, there is a significant difference in efficiency between both protocols.

\begin{figure}
\centering
\includegraphics[width=0.7\linewidth]{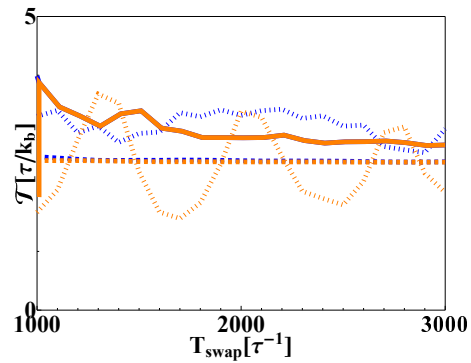}
\caption{ Comparison of the different measures of temperature as a function of $T_{\text{swap}}$ for the (0.01-shifted) tight-binding dispersion. Dotted lines: $\mathcal{T}_{\text{ext}}$, dashed lines: $\mathcal{T}_{BG}$ and full lines: $\mathcal{T}_{\text{av}}$. The blue lines correspond to protocol 1 and the orange ones to protocol 2.}
\label{fig:tbtemperatures}
\end{figure}
\FloatBarrier

\bibliographystyle{apsrev4-1}
%

\end{document}